\documentclass[conference]{IEEEtran}
\IEEEoverridecommandlockouts
\usepackage{verbatim}
\usepackage{cite}
\usepackage{amsmath,amssymb,amsfonts}
\usepackage{algorithmic}
\usepackage{graphicx}
\usepackage{textcomp}
\usepackage{xcolor}
\usepackage{tikz}
\usepackage{environ}
\makeatletter
\newsavebox{\measure@tikzpicture}
\NewEnviron{scaletikzpicturetowidth}[1]{%
  \def\tikz@width{#1}%
  \begin{lrbox}{\measure@tikzpicture}%
  \BODY
  \end{lrbox}%
  \pgfmathparse{#1/\wd\measure@tikzpicture}%
  \BODY
}
\makeatother
\usepackage{pgfplots}
\usepackage{tikz}
\usetikzlibrary{calc}

\usepackage[nocomma]{optidef}
\newtheorem{theorem}{Theorem}
\newtheorem{corollary}{Corollary}

\begin{document}

\title{Holographic MIMO: How Many Antennas Do We Need for Energy Efficient Transmission?
}

\author{\IEEEauthorblockN{Sarah Bahanshal, Qurrat-Ul-Ain Nadeem, and Md. Jahangir Hossain}
\IEEEauthorblockA{School of Engineering, The University of British Columbia, Kelowna, Canada}
Email: \{sarah.bahanshal, qurrat.nadeem, jahangir.hossain\}@ubc.ca
}

\maketitle

\begin{abstract}
 Holographic multiple-input multiple-output (HMIMO) communication systems utilize spatially-constrained massive MIMO arrays containing large numbers of antennas with sub-wavelength spacing, and have emerged as a promising candidate technology for Sixth Generation (6G) networks. In this paper, we consider the downlink of a multi-user HMIMO communication system under a Fourier plane-wave series representation of the stochastic electromagnetic MIMO channel model, and make two important contributions. First, we present a closed-form expression of the ergodic achievable downlink rate under maximum ratio transmission (MRT) precoding at the base station (BS). The derived expression explicitly shows the effect of the side-lengths of the HMIMO surfaces at the BS and each user, and the number of antennas deployed in these surfaces on the user rates. Second, we formulate an energy efficiency (EE) maximization problem with respect to the number of antennas arranged within spatially-constrained HMIMO surfaces at the BS and each user. The resulting implicit solution for this problem is shown to be globally optimal. Numerical results yield useful insights into the EE performance of multi-user HMIMO systems in different operating regimes.
\end{abstract}


\section{Introduction}

The BSs in  Fifth Generation (5G) cellular networks are equipped with a large number of antennas to enable directed beamforming towards multiple users, resulting in what is known as massive  MIMO systems \cite{ref_mmimo, ref_howmanyant,ref_mimogen}. The spectral efficiency of massive MIMO systems is shown to increase significantly with the number of antennas under  the assumption that the array aperture also grows large with the number of antennas \cite{ref_howmanyant}. 
As a result, massive MIMO arrays with upto $128$ antennas are considered to meet the 5G wireless networks' throughput needs.
However,  the data rate requirements are becoming more stringent  as research focus shifts towards 6G communication systems. Further increasing the number of antennas  at the communication end-points to meet these requirements is practically challenging due to space-limitations at the BS and user equipment \cite{ref_mmimo,ref_channelmodel2}.

 In light of these challenges, HMIMO communication has emerged as a candidate technology for 6G networks, and refers to communication between spatially-constrained arrays equipped with a massive number of densely deployed antennas \cite{ref_surv50hmimo}. In the asymptotic limit, the arrays represent spatially-continuous electromagnetic apertures that can actively generate and receive beamformed radio signals and can fully exploit the propagation characteristics offered by an electromagnetic channel \cite{ref_hmimochannelmodel,ref_mmimo,ref_hmimotrends}. Since the antennas in HMIMO surfaces are closely spaced with less than half-a-wavelength spacing, a new channel model has been developed that captures the electromagnetic propagation between  these surfaces while accounting for the inherent spatial correlation \cite{ref_hmimochannelmodel,ref_channelmodel2}.
 

In \cite{ref_hmimochannelmodel,ref_channelmodel2}, a Fourier plane-wave series representation of the channel response between any two points on the transmit and receive electromagnetic surfaces is introduced. This spatially-continuous channel model fully captures the essence of electromagnetic propagation  under arbitrary scattering and is  valid in both near-field and far-field propagation regions. The derived channel representation is then sampled to obtain a stochastic electromagnetic MIMO channel matrix for a point-to-point HMIMO system in \cite{ref_hmimochannelmodel,ref_channelmodel2} and for a multi-user HMIMO system in \cite{ref_muhmimo}. Research on HMIMO systems is still in its infancy, and theoretically analyzing  the performance under spatially-continuous electromagnetic channel models is difficult. Some works have investigated the number of spatial degrees-of-freedom (DoF) these systems offer \cite{ref_dofhmimo,ref_hmimocommodes}. More recently, the authors in  \cite{ref_muhmimo} studied the spectral efficiency of a multi-user HMIMO system under different precoding schemes. In \cite{ref_irselements}, the number of antennas at the BS and the number of intelligent reflecting surface elements were optimized to maximize the EE. 

In this work, we consider the downlink multi-user HMIMO communication model under a stochastic electromagnetic MIMO channel representation \cite{ref_hmimochannelmodel}. Then, we provide a closed-form expression for the ergodic achievable downlink rate at each user under MRT precoding at the BS. The expression is a function of the side lengths of the HMIMO surfaces,   the number of antennas in these surfaces, and the signal-to-noise ratio (SNR). Next we study  the number of antennas that should be placed within the spatially constrained arrays at the BS and the users such that the EE of the HMIMO communication system is maximized. To do this, we formulate an EE maximization problem using the derived achievable rate expression and a practical HMIMO power consumption model, and solve the problem analytically. Numerical results reveal interesting insights into the performance of HMIMO systems in different operating regimes.

\section{System Model}
We consider the downlink communication  between a BS and $K$ users, all equipped with  planar HMIMO surfaces made up of a large number of sub-wavelength spaced patch antennas \cite{ref_hmimochannelmodel,ref_mmimo}. The HMIMO surface at the BS has finite side lengths $L_{s,x}$ and $L_{s,y}$ and is equipped with $N_s=N_{H_s}N_{V_s}$ patch antennas, where $N_H$ and $N_V$ represent the number of horizontally and vertically arranged patch antennas with inter-antenna spacing $\Delta_{s,x}$ and 
$\Delta_{s,y}$ units, respectively. The patch antennas at the BS's HMIMO surface are indexed by $n= 1,\ldots,N_s$, so as the position vector of the $n^\text{th}$ patch antenna with respect to the origin is 
$\mathbf{s}_n=[s_{x,n},s_{y,n},s_{z,n}]^T=[i(n)\Delta_{s,x},j(n)\Delta_{s,y},0]^T$, where $i(n)=\mod(n-1,N_{H_s})$, and $j(n)=\lfloor{(n-1)/N_{H_s})}\rfloor$. Similarly, each of the $K$ users' HMIMO surface has finite lengths $L_{r,x}$ and $L_{r,y}$ and is equipped with $N_r=N_{H_r} N_{V_r}$ antennas spaced by $\Delta_{r,x}$ and $\Delta_{r,y}$ units in the horizontal and vertical directions respectively. The position of the $m^\text{th}$ antenna of the $k^\text{th}$ user is $\mathbf{r}_m^k=[r_{x,m}^k,r_{y,m}^k,r_{z,m}^k]^T=[i(m)\Delta_{r,x},j(m)\Delta_{r,y},d_k]^T$, where $d_k$ is the $z$ coordinate of the $k^\text{th}$ user.

We model the channel between   two arbitrary points   $\mathbf{r}^k$ and $\mathbf{s}$ at user $k$'s and BS's HMIMO surfaces respectively using the Fourier plane wave series expansion of the spatial impulse response $h(\mathbf{r}^k,\mathbf{s})$ outlined in \cite{ref_hmimochannelmodel,ref_channelmodel2}. Sampling the channel impulse response $h(\mathbf{r}^k,\mathbf{s})$ at points $\mathbf{s}_n$ and $\mathbf{r}_m^k$ for $n=1,\dots, N_s$ and $m=1,\dots, N_r$ yields the stochastic HMIMO channel matrix that is outlined next. To this end, the transmit wave vector at the BS's HMIMO surface is denoted by $\mathbf{u}_s(m_x,m_y)\in \mathbb{C}^{N_s}$ with entries $\frac{1}{\sqrt{N_s}}a_s(m_x,m_y,\mathbf{s}_n)$ for $n=1,\ldots,N_s$, where $a_s(m_x,m_y,\mathbf{s}_n)$ represent the discretized plane-wave harmonics and are defined as \cite{ref_hmimochannelmodel}
\begin{align}
\label{Txh}
a_s(m_x,m_y,\mathbf{s}_n)= e^{j\left(\frac{2\pi}{L_{s,x}}m_x s_{x,n}+\frac{2\pi}{L_{s,y}}m_y s_{y,n}+\gamma_s(m_x,m_y)s_{z,n}\right)},
\end{align}
where $\gamma_s(m_x,m_y)=\sqrt{\kappa^2-\left(\frac{2\pi m_x}{L_{s,x}}\right)^2-\left(\frac{2\pi m_y}{L_{s,y}}\right)^2}$, $\kappa=2\pi /\lambda$, and $\lambda$ is the wavelength. Similarly, $\mathbf{u}_r^k(\ell_x,\ell_y)\in \mathbb{C}^{N_r}$ is the receive wave vector for user $k$ with entries $\frac{1}{\sqrt{N_r}}a_r(\ell_x,\ell_y,\mathbf{r}_m^k)$ for $m=1,\ldots,N_r$ defined as \cite{ref_hmimochannelmodel}
\begin{align}
\label{Rxh}
a_r(\ell_x,\ell_y,\mathbf{r}_m^k)= e^{j\left(\frac{2\pi}{L_{r,x}}\ell_x r_{x,m}^k+\frac{2\pi}{L_{r,y}}\ell_y r_{y,m}^k+\gamma_r(\ell_x,\ell_y)r_{z,m}^k\right)},
\end{align}
where $\gamma_r(\ell_x,\ell_y)$ is defined similar to $\gamma_s(m_x,m_y)$. In \eqref{Txh} and \eqref{Rxh}, $(m_x,m_y)$ and $(\ell_x, \ell_y)$ represent the integer coordinates of the lattice ellipses in which the plane-wave harmonics are non-zero, with the lattice ellipse defined as  
$\xi_s=\{(m_x,m_y)\in \mathbb{Z}^2: \left(m_x\lambda/L_{s,x}\right)^2+\left(m_y\lambda/L_{s,y}\right)^2\leq 1\}$ and $\xi_r=\{(\ell_x,\ell_y)\in \mathbb{Z}^2: \left(\ell_x\lambda/L_{r,x}\right)^2+\left(\ell_y\lambda/L_{r,y}\right)^2\leq 1\}$ at the BS and user respectively \cite{ref_hmimochannelmodel}. We denote  the cardinalities of $\xi_s$ and $\xi_r$ as $n_s=|\xi_s|$ and $n_r=|\xi_r|$ respectively. It was shown in \cite{ref_channelmodel2,ref_hmimochannelmodel} that for $\text{min}(L_{s,x},L_{s,y})/\lambda \gg 1$   and $\text{min}(L_{r,x},L_{r,y})/\lambda \gg 1$
\begin{align}
\label{eq_n_sr}
n_s\approx\left\lfloor\frac{\pi}{\lambda^2}L_{s,x}L_{s,y}\right\rfloor, \;\;\;
n_r\approx\left\lfloor\frac{\pi}{\lambda^2}L_{r,x}L_{r,y}\right\rfloor,
\end{align}
where $n_s$ and $n_r$ quantify the number of DoF at the transmitter and receiver respectively. To guarantee that no information is lost by the spatial sampling of $h(\mathbf{r}^k, \mathbf{s})$ at points $\mathbf{s}_n$ and $\mathbf{r}_m^k$, the Nyquist condition in the spatial domain must be satisfied and requires that \cite{ref_channelmodel2,ref_hmimochannelmodel} 
\begin{align}
\label{Ncons}
&N_s\geq n_s,\text{ and }  N_r\geq n_r.
\end{align}

Utilizing these definitions, the HMIMO channel matrix $\mathbf{H}^k \in \mathbb{C}^{N_r\times N_s}$ between the BS and $k^{\text{th}}$ user is given by
\begin{align}
\label{fchannel}
\mathbf{H}^k=\mathbf{U}_r^k\mathbf{H}_a^k\mathbf{U}_s^H,  
\end{align}
where $\mathbf U_s\in \mathbb C^{N_s\times n_s}$ collects the $n_s$ transmit column vectors $\mathbf u_s(m_x,m_y)$ and $\mathbf U_r^k\in \mathbb C^{N_r\times n_r}$ collects the $n_r$ receive column vectors $\mathbf u_r^k(\ell_x,\ell_y)$ for user $k$, such that $\mathbf U_s^H \mathbf U_s=\mathbf I_{n_s}$ and ${\mathbf U_r^k}^H \mathbf U_r^k=\mathbf I_{n_r}$. Moreover, $\mathbf H_a^k\in \mathbb C^{n_r\times n_s}$ is the wavenumber domain channel between BS and user $k$, and represents the angular response that maps the transmit directions to the receive directions. According to \cite{ref_hmimochannelmodel}, $\mathbf H_a^k$ collects the zero mean independent random variables $\sqrt{N_rN_s}H_a^k(\ell_x,\ell_y,m_x,m_y)$, where each variable follows the complex Gaussian distribution as
\begin{align}
\label{fchannel1}
&H_a^k(\ell_x,\ell_y,m_x,m_y)\sim \mathcal{N}_{\mathbb C}(0,\sigma^{2,k}(\ell_x,\ell_y,m_x,m_y)),
\end{align}
where $\sigma^{2,k}(\ell_x,\ell_y,m_x,m_y)$ represents the fraction of power transferred between the angular sets (discretized wavenumbers) $(m_x,m_y)$ and $(\ell_x,\ell_y)$ at the transmit and receive sides respectively.  For simplicity, we assume that the scattering propagation scenario has variance separability \cite{ref_hmimochannelmodel,ref_channelmodel2}, i.e., $\sigma^{2,k}(\ell_x,\ell_y,m_x,m_y)=\sigma_r^{2,k}(\ell_x,\ell_y)\sigma_s^2(m_x,m_y)$. 
To this end, we denote the $n_r$ non-zero $\sigma_r^{2,k}(\ell_x,\ell_y)$ as ${\sigma_{r,i}^{2,k}}$ for $i=1,\ldots,n_r$, while the $n_s$ non-zero $\sigma_s^2(m_x,m_y)$ are denoted as ${\sigma_{s,t}^2}$ for $t=1,\ldots,n_s$. The variances can be computed using \cite[(34)]{ref_hmimochannelmodel}. Note that the DoF offered by the HMIMO channel $\mathbf{H}^k$ in \eqref{fchannel} is $\min(n_s, n_r)$.

Under this electromagnetic HMIMO channel model, the received signal at all the users, $\mathbf{y}\in\mathbb{C}^{N_rK\times 1}$, is given as
\begin{equation}
    \mathbf y=\sqrt{p_u} \mathbf H \mathbf V \mathbf{x}+\mathbf{w}=\sqrt{p_u} \mathbf U_r\mathbf H_a \mathbf U_s^H \mathbf V \mathbf{x}+\mathbf{w},
\end{equation}
where  $\mathbf{H}\in \mathbb{C}^{N_rK\times N_s}$ is the concatenated channel between the BS and all users, i.e., $\mathbf{H}=\left[{\mathbf{H}^1}^T {\mathbf{H}^2}^T\ldots {\mathbf{H}^K}^T\right]^T
$, $\mathbf{V}\in \mathbb{C}^{N_s\times n_rK}$ is the precoding matrix at the BS, $\mathbf{x}\in\mathbb{C}^{n_rK\times 1}$ is the transmit signal vector for all users, $p_u$ is the transmit power, and $\mathbf{w}\in\mathbb{C}^{N_rK\times 1}$ is the additive Gaussian noise with i.i.d. elements having zero mean and variance $\sigma_w^2$.  

\section{Downlink Achievable Rates}
This section derives the ergodic achievable downlink rates under MRT precoding and the HMIMO channel model in \eqref{fchannel}.

\subsection{Achievable Rate Formulation}

Following \cite{ref_muhmimo}, let $\mathbf{y}_a=\mathbf{U}_r^H\mathbf{y}\in \mathbb{C}^{n_rK\times 1}$ be the received signal in the wavenumber domain. Defining $\tilde{\mathbf{H}}_a=\mathbf{H}_a\mathbf{U}_s^H\in \mathbb{C}^{n_rK\times N_s}$ and $\mathbf w_a=\mathbf{U}_r^H\mathbf w$,  we can write
\begin{align}
\label{fchannel2}
    \mathbf y_a=\sqrt{p_u}\tilde{\mathbf{H}}_a\mathbf V \mathbf{x}+\mathbf{w}_a.
\end{align}
Using \eqref{fchannel2}, the  signal received at user $k$ corresponding to the $i^{\text{th}}$  sampling point (or DoF), where $i=1,\dots, n_r$, is given as
\begin{align}
\label{eqn_decyperdof1}
y_{a,i}^{k}&=\sqrt{p_u}{\mathbf{g}_i^k}^H\mathbf{v}_i^k{x}_{i}^k+\sqrt{p_u}\sum_{j\neq i}^{n_r}\sum_{q\neq k}^{K}{\mathbf{g}_i^k}^H\mathbf{v}_j^q x_{j}^q+w_{a,i}^k,
\end{align}
where ${\mathbf{g}_i^k}^H\in \mathbb{C}^{1\times{N_s}}$ is the $i^\text{th}$ row of the $k^{\text{th}}$ sub-block of ${\tilde{\mathbf{H}}}_a$ and represents the channel between the BS and user $k$ corresponding to the $i^{\text{th}}$ receive sampling point. Also, $\mathbf{v}_i^k \in \mathbb{C}^{N_s\times 1}$ is the $i^{\text{th}}$ column of the  $k^{\text{th}}$ sub-block of  $\mathbf{V}$, and $w^k_{a,i}$ is the $i^\text{th}$ element of the $k^\text{th}$ sub-column of $\mathbf{w}_a$.

For the received signal model above, we present an ergodic achievable rate expression for each user, exploiting a technique from \cite{ref_ach_rate}, which is widely applied in works on massive MIMO systems \cite{ref_howmanyant}. The technique exploits the channel hardening property of massive MIMO systems, as $N_s$ grows large, the effective channel ${\mathbf{g}_i^k}^H{\mathbf{v}_i^k}$  at user $k$ approaches its average value $\mathbb{E}\left[{\mathbf{g}_i^k}^H\mathbf{v}_i^k\right]$. Under this property, the authors  assume the availability of only channel statistics (i.e. knowledge of $\mathbb{E}\left[{\mathbf{g}_i^k}^H\mathbf{v}_i^k\right]$) at the users to compute the SINR. The main idea then is to decompose $y_{a,i}^k$ in \eqref{eqn_decyperdof1} for $i=1,\ldots,n_r$ as


\begin{align}
\label{eqn_decyperdof}
y_{a,i}^k&=\sqrt{p_u}\mathbb{E}\left[{\mathbf{g}_i^k}^H\mathbf{v}_i^k\right]x_{i}^k+\sqrt{p_u}\left(\mathbf{g}_i^H\mathbf{v}_i-\mathbb{E}\left[{\mathbf{g}_i^k}^H\mathbf{v}_i^k\right]\right){x}_{i}\nonumber \\
&+\sqrt{p_u}\sum_{j\neq i}^{n_r}\sum_{q\neq k}^{K}{\mathbf{g}_i^k}^H\mathbf{v}_j^q x_{j}^q+w^k_{a,i},
\end{align}
and assume that the average effective channel $\mathbb{E}\left[{\mathbf{g}_i^k}^H\mathbf{v}_i^k\right]$ is perfectly known at user $k$. Using  \eqref{eqn_decyperdof} and  treating interference and channel uncertainty as worst-case independent Gaussian noise, we conclude that user $k$ can achieve the following ergodic rate and signal-to-interference-plus-noise ratio (SINR) corresponding to the $i\text{th}$ receive DoF \cite[Thm. 1]{ref_ach_rate}
\begin{align}
\label{rate}
R_i^k&=\log_2(1+\text{SINR}_i^k),\\
\label{SINR}
    \text{SINR}_i^k&=\frac{p_u \left|\mathbb{E}\left[\mathbf{g}_i^{k^H}\mathbf{v}_i^k\right]\right|^2}{\sigma_w^2+p_u \text{Var}[\mathbf{g}_i^{k^H}\mathbf{v}_i^k]+p_u \sum_{j\neq i}^{n_r}\sum_{q\neq k}^{K}\mathbb{E}[|\mathbf{g}_i^{k^H}\mathbf{v}_j^q|^2]},
\end{align}
where $ \text{Var}[x]$ denotes the variance of  $x$. The ergodic achievable sum rate of the  HMIMO system is then given as 
\begin{align}
\label{R_sum}
&\mathcal{R}_\text{sum}=\sum_{k=1}^{K} \sum_{i=1}^{n_r}  \mathcal{R}_i^k= \sum_{k=1}^{K} \sum_{i=1}^{n_r}\log_2\left(1+\text{SINR}_i^k\right).
\end{align}

\subsection{Achievable Rate Expression under MRT}
For MRT precoding, the precoding matrix is given as 
\begin{align}
&\mathbf V= \alpha_\text{MRT}\Tilde{\mathbf{H}}_a^H,
\end{align}
where $\alpha_{\text{MRT}}$ is a normalization coefficient to ensure that the average transmit power constraint, $\mathbb{E}\left[\text{Tr}(\mathbf{V}\mathbf{V}^H)\right]=1$ is satisfied. Using $\mathbf{U}_s^H\mathbf{U}_s=\mathbf{I}_{n_s}$, we obtain
\begin{align}
    \alpha_{\text{MRT}}^2&={\frac{1}{\mathbb{E}[\text{Tr}(\Tilde{\mathbf{H}}_a\Tilde{\mathbf{H}}_a^H)]}}
    ={\frac{1}{\mathbb{E}\left[\text{Tr}({\mathbf{H}}_a{\mathbf{H}}_a^H)\right]}}.
\end{align}

The SINR in \eqref{SINR}  under MRT precoding is given as $\text{SINR}_i^{k,\text{MRT}}=$\vspace{-.15in}
\begin{align}
\label{eqn_SINR_MRT}
&\frac{p_u \alpha_{\text{MRT}}^2|\mathbb{E}[\mathbf{g}_i^{k^H}\mathbf{g}_i^k]|^2}{\sigma_w^2+p_u \alpha_{\text{MRT}}^2\text{Var}[\mathbf{g}_i^{k^H}\mathbf{g}_i^k]+p_u \alpha_{\text{MRT}}^2\sum_{j\neq i}^{n_r}\sum_{q\neq k}^{K}\mathbb{E}[\mathbf{g}_i^{k^H}\mathbf{g}_j^q|^2]}
\end{align}

To obtain a closed-form expression of ergodic achievable rate, we compute the expected value and variance terms in \eqref{eqn_SINR_MRT} using the channel model in \eqref{fchannel} and exploiting results on the statistics of Gaussian random variables as follows:
\begin{align}
    \label{eqn_alpha_mrt2}
\alpha_{\text{MRT}}^2&={\frac{1}{N_rN_s\sum_{k=1}^{K}\sum_{i=1}^{n_r}\sum_{t=1}^{n_s}\sigma_{r,i}^{2,k}\sigma_{s,t}^2}}, \\
    \mathbb{E}\left[\left|{\mathbf{g}_i^k}^H \mathbf g_{j}^q\right|^2\right]&=\left(\sigma_{r,i}^k\sigma_{r,j}^qN_rN_s\right)^2\sum_{t=1}^{n_s}\sigma_{s,t}^4, \\
    \mathbb{E}\left[{\mathbf{g}_i^k}^H\mathbf{g}_i^k\right]&=N_rN_s\sigma_{r,i}^{2,k}\sum_{t=1}^{n_s}\sigma_{s,t}^2, \\
    \text{Var}\left[{\mathbf{g}_i^k}^H\mathbf{g}_i^k\right]&=(N_r N_s)^2\sigma_{r,i}^
    {4,k} \sum_{t=1}^{n_s} \sigma_{s,t}^4.
    \label{eqn_var_gigi}
\end{align}

Utilizing these results, the ergodic achievable rate at user $k$  can be written compactly as given in the following theorem.
\begin{theorem}
The ergodic achievable rate of user $k$ corresponding to  $i=1, \dots, n_r$ receive DoFs under  MRT precoding is given as \vspace{-.1in}
\begin{align}
\label{Th1}
\mathcal{R}_i^{k,\text{MRT}}=\log\left(1+\frac{\sigma_{r,i}^{2,k}\left(n_s\hat{\sigma}_s^2\right)^2}{\frac{\sigma_w^2}{\sigma_{r,i}^{2,k}p_u \alpha_{\text{MRT}}^2(N_r N_s)^2}+n_r\hat{\sigma}_r^2\sum_{t=1}^{n_s}\sigma_{s,t}^4}\right),
\end{align} 
where $\hat{\sigma}_{s}^2=\frac{1}{n_s}\sum_{t=1}^{n_s}\sigma_{s,t}^2$ and $\hat{\sigma}_{r}^2=\frac{1}{n_r}\sum_{i=1}^{n_r}\sum_{k=1}^{K}\sigma_{r,i}^{2,k}$
\end{theorem} 
\begin{IEEEproof}
Using \eqref{eqn_alpha_mrt2}-\eqref{eqn_var_gigi} in \eqref{eqn_SINR_MRT} yields \eqref{eqn_rate_mrt_step} (given at the top of next page), which is simplified and substituted in \eqref{rate} to obtain \eqref{Th1}.
\end{IEEEproof}
\begin{figure*}[t]
\begin{align}
\label{eqn_rate_mrt_step}
\mathcal{R}_i^{k,\text{MRT}}=\log \left(1+\frac{p_u \alpha_{\text{MRT}}^2\left(N_rN_s\sigma_{r,i}^{2,k}\sum_{t=1}^{n_s}\sigma_{s,t}^2\right)^2}{\sigma_w^2+p_u \alpha_{\text{MRT}}^2(N_r N_s)^2\sigma_{r,i}^{4,k}\sum_{t=1}^{n_s} \sigma_{s,t}^4+p_u \alpha_{\text{MRT}}^2\sigma_{r,i}^{2,k}(N_r N_s)^2\sum_{t=1}^{n_s}\sigma_{s,t}^4\sum_{j\neq i}^{n_r}\sum_{q\neq K}^{K}\sigma_{r,j}^{2,q}}\right).
\end{align}
\end{figure*}

\begin{figure}[t]
\centering
\includegraphics[width=.85\columnwidth]{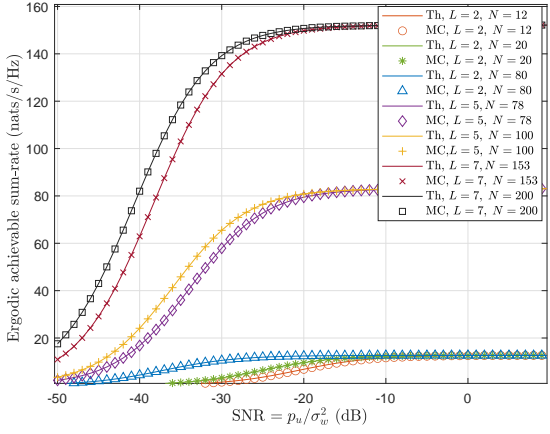}
\caption{Ergodic achievable sum rate for $K=3$, $L_{s,x}=L_{s,y}=L\lambda$ and  $L_{r,x}=L_{r,y}=L \lambda.$ Both theoretical (Th) and Monte-Carlo (MC) simulated sum-rate are plotted for different number of antennas $N=N_s=N_r$.}
\label{fig_rate}
\end{figure}

The expression in Theorem 1 depends explicitly on the number of antennas constituting the HMIMO surfaces at the BS and user ends, i.e. $N_s$ and $N_r$ respectively, the number of users $K$, and the number of DoF created by the scattering at the transmit and receive sides, i.e. $n_s$ and $n_r$ respectively, which are a function of $L_{s,x}$, $L_{s,y}$, $L_{r,x}$, $L_{r,y}$ and $\lambda$. The first term in the denominator of \eqref{Th1} represents the noise, and the second term represents the effect of multi-user interference. Interestingly, we see that increasing $N_s$ and $N_r$ beyond $n_s$ and $n_r$ respectively decreases the noise but does not impact the interference. It is therefore important to study how many additional antennas are needed beyond the DoF offered by the  channel to yield energy-efficient HMIMO systems.

To validate the derived expression in Theorem 1, we plot in Fig. \ref{fig_rate} the sum-rate in \eqref{R_sum} using the Monte-Carlo simulated SINR  in \eqref{eqn_SINR_MRT} as well the theoretical SINR in \eqref{Th1}. The figure is plotted for  $K=3$,  $L_{s,x}=L_{s,y}=L\lambda$, and $L_{r,x}=L_{r,y}=L\lambda$, for different values of $L$ and $N=N_s=N_r$.  The curves show a very close match between the theoretical expression and Monte Carlo simulated sum-rate with 1000 iterations. The figure also shows that as the side lengths of the source and receive HMIMO surfaces increase, the sum-rate increases. This is because the DoF offered by the HMIMO matrix is given by $\text{min}(n_s, n_r)$, and $n_s$ and $n_r$ increase with the side lengths as shown in \eqref{eq_n_sr}. Furthermore, we see that if we keep the lengths $L_{s,x}=L_{s,y}=L_{r,x}=L_{r,y}$ fixed, then adding more antennas within the  spatially constrained HMIMO surfaces is only beneficial at low to moderate SNR values. For example, when $L_{s,x}=L_{s,y}=L_{r,x}=L_{r,y}=2\lambda$  (which results   in $n_r=n_s=12$ DoF using \eqref{eq_n_sr}), we can see that for low and moderate SNR values the  sum-rate is higher when $N>>12$ antennas are placed in the HMIMO surfaces, whereas  for high SNR values the sum-rate is the same for $N=12$ and $N=80$ antennas, and having additional antennas above the limits in \eqref{Ncons} does not increase the sum-rate.  

\section{Energy Efficiency Optimization}
In this section, we outline the EE of the considered HMIMO  system, and maximize it with respect to $N_s$ and $N_r$. 

\subsection{Power Consumption Model}
Consider the power consumption model of HMIMO surfaces proposed in \cite{ref_energy_hmimos}. According to this model, the total power consumption of the HMIMO surface at the BS is 
\begin{align}
P_s=N_sL_dP_d+\frac{N_s}{Q}P_v+P_f+\frac{1}{\zeta}p_u,
\end{align}
where $L_d$ is the number of diodes per holographic antenna patch, $Q$ is the number of holographic antenna patches per group, $P_d, P_v, P_f$ are the power consumption of one diode, one voltage converter and one FPGA respectively in watts (W), and $\zeta$ is the power amplifier efficiency  \cite{ref_energy_hmimos}.

Similarly, the power consumption at user $k$'s surface is 
\begin{align}
P_r=N_rL_dP_d+\frac{N_r}{Q}P_v+P_f,
\end{align}
and the total power consumption of the HMIMO system is 
\begin{align}
    P_{\text{tot}}&=P_s+KP_r=(N_s+KN_r)P_1+P_2,
\end{align}
where $P_1=L_dP_d+P_v/Q$ and $P_2=(K+1)P_f+\frac{1}{\zeta}p_u$. 
\subsection{Energy Efficiency Problem Formulation}
The EE (in nats per Hz per joule) is  defined as the ratio of the  ergodic achievable sum-rate and the total power consumption of the HMIMO communication system as
\begin{align}      \text{ EE}=\frac{\sum_{i=1}^{n_r}\sum_{K=1}^{K}\mathcal{R}_i^{k,\text{MRT}}}{(N_s+KN_r)P_1+P_2}.
\end{align}
To this end, we formulate an EE maximization problem to find the optimal numbers of antennas $N_s$ and $N_r$  to be placed within given BS and users' HMIMO surface side lengths ($L_{s,x},L_{s,y}$) and ($L_{r,x},$ and $L_{r,y}$) respectively. Note that by fixing $L_{s,x},L_{s,y},L_{r,x},$ and $L_{r,y}$, the number of transmit and receive DoFs $n_s$ and $n_r$ are fixed, and we focus on the impact of having antennas in excess of the DoF on the EE. The EE maximization problem \textit{(P1)}  is formulated next.
\begin{subequations}
\begin{align}\label{eqn_maxee}
\textit{(P1)} &\hspace{.2in} \max_{N_r,N_s} \frac{\sum_{i=1}^{n_r}\sum_{k=1}^{K}\log\left(1+\frac{a_i^k}{\frac{b_i^k}{N_rN_s}+c}\right)}{(N_s+KN_r)P_1+P_2}\\
	\rm \;\;\;\;s.t. \label{C1}
	&	\hspace{0.2cm} \hspace{0.2cm} N_r\geq  n_r, \\ \label{C2}
	&\hspace{0.4cm} N_s\geq  n_s,	\end{align}
\end{subequations}
 where $a_i^k={\sigma_{r,i}^k}^2\left(n_s\hat{\sigma}_s^2\right)^2$, $b_i^k=\frac{\sigma_w^2\sum_{k=1}^{K}\sum_{i=1}^{n_r}\sum_{t=1}^{n_s}{\sigma_{r,i}^k}^2\sigma_{s,t}^2}{{\sigma_{r,i}^k}^2p_u}$, and $c=n_r\hat{\sigma}_r^2\sum_{t=1}^{n_s}\sigma_{s,t}^4$. Constraints \eqref{C1} and \eqref{C2} stem from \eqref{Ncons}, and ensure that by sampling the spatially continuous channel impulse response at $N_s$ and $N_r$ points, the spatial-Nyquist condition is satisfied and no information is lost  \cite{ref_hmimochannelmodel}.

\subsection{Optimization Problem Solution} \label{sec_optboth}
This section presents the optimal solution to \textit{(P1)}. We use Karush–Kuhn–Tucker (KKT) conditions to find the optimal number of antennas to be placed in the BS and users' HMIMO surfaces, denoted by $N_s^*$ and $N_r^*$ respectively, that maximize the EE while satisfying the constraints \eqref{C1} and \eqref{C2}. First,  the Lagrangian function of \textit{(P1)} is expressed as
\begin{align}
\begin{split}
L(N_s,&N_r,\mu_1,\mu_2)=-\mu_1(n_s-N_s)-\mu_2(n_r-N_r)\\
&+\frac{\sum_{k=1}^{K}\sum_{i=1}^{n_r}\log\left(1+\frac{a_i^k}{\frac{b_i^k}{N_rN_s}+c}\right)}{(N_s+KN_r)P_1+P_2},
\end{split}
\end{align}
where $\mu_1$, and $\mu_2$ are the Lagrange multipliers associated with the inequality constraints \eqref{C1} and \eqref{C2}, respectively. The KKT conditions for this problem are given as $N_s^*\geq n_s, N_r^*\geq n_r, \mu_1^*\geq 0, \mu_2^*\geq 0, \mu_1^*(n_s-N_s^*)=0, \mu_2^*(n_r-N_r^*)=0, \frac{\partial}{\partial N_s}L(N_s^*,N_r^*,\mu_1^*,\mu_2^*)=0,$ and $\frac{\partial}{\partial N_r}L(N_s^*,N_r^*,\mu_1^*,\mu_2^*)=0$. The solution after solving these conditions is presented next.

\begin{figure}[t]
\centering
\includegraphics[width=.85\columnwidth]{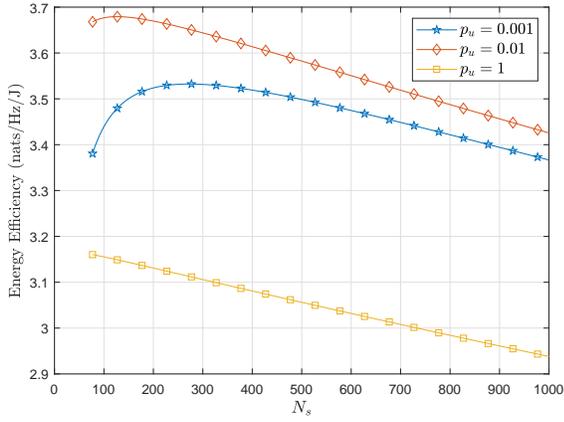}
\caption{EE versus the number of source antennas $N_s$ for different $p_u$ values, $K=3$ users, $L_{s,x}=L_{s,y}=5\lambda$, $L_{r,x}=L_{r,y}=1\lambda$, and fixed $N_r=N_r^*$.}
\label{fig_ee1}
\end{figure}

\begin{theorem} The optimal numbers of antennas $N_s^*$ and $N_r^*$ at the BS and each user respectively, that satisfy all the aforementioned KKT conditions can be obtained as
\begin{align}
&(N_s^*, N_r^*)=\begin{cases}
(\bar{N_s}, \bar{N_r}) & \text{ for } \bar{N}_s>n_s  \text{ and } \bar{N}_r>n_r \\
(\bar{N_s}(n_r), n_r)  & \text{ for } \bar{N}_s>n_s \text{ and } \bar{N}_r\leq n_r \\
(n_s, \bar{N_r}(n_s))  & \text{ for } \bar{N}_s\leq n_s \text{ and } \bar{N}_r> n_r \\
(n_s, n_r) & \text{ for } \bar{N}_s\leq n_s \text{ and } \bar{N}_r \leq n_r 
\end{cases}
\end{align}
where $\bar{N_s}$ and $\bar{N_r}$ are obtained as the solution of  \eqref{eq_implicitX} and \eqref{eq_implicity}, $\bar{N_s}(n_r)$ is obtained  by solving \eqref{eq_implicitX} for $\bar{N}_s$ by plugging $\bar{N}_r=n_r$, and $\bar{N_r}(n_s)$ is obtained  by solving \eqref{eq_implicity}  for $\bar{N}_r$ by plugging $\bar{N}_s=n_s$. Note that the values of $\bar{N_s}$ and $\bar{N_r}$ are  rounded to the closest integers that give the higher EE.


\end{theorem}
\begin{IEEEproof}
    The proof involves solving the aforementioned KKT conditions and is omitted for brevity in writing.
\end{IEEEproof}
\begin{corollary}
The solution ($N_s^*, N_r^*$) in Theorem 2 for \textit{(P1)} is unique under constraints \eqref{C1} and \eqref{C2}. 
\end{corollary}
\begin{IEEEproof}
The proof is given in the appendix.
\end{IEEEproof}
\begin{figure}[t]
\centering
\includegraphics[width=.85\columnwidth]{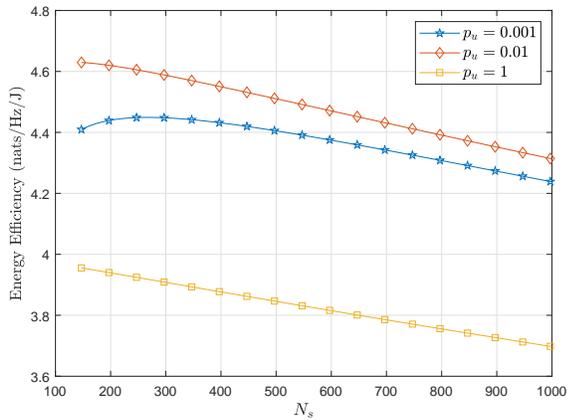}
\caption{EE versus the number of source antennas $N_s$ for different $p_u$ values, $K=3$ users, $L_{s,x}=L_{s,y}=7\lambda$, $L_{r,x}=L_{r,y}=1\lambda$, and fixed $N_r=N_r^*$.}
\label{fig_ee2}
\end{figure}
\begin{figure*}[]\small
\begin{align}
\label{eq_implicitX}
\left(P_1(\bar{N_s}+K\bar{N_r})+P_2\right)\left(\sum_{i=1}^{n_r}\sum_{k=1}^{K}\frac{a_i^kb_i^k\bar{N_r}}{(c\bar{N_r}\bar{N_s}+b_i^k)((c+a_i^k)\bar{N_r}\bar{N_s}+b_i^k)}\right)-
P_1\left(\sum_{i=1}^{n_r}\sum_{k=1}^{K}\log\left(1+\frac{a_i^k}{\frac{b_i^k}{\bar{N_r}\bar{N_s}}+c}\right)\right)=0\\
\label{eq_implicity}
\left(P_1(\bar{N_s}+K\bar{N_r})+P_2\right)\left(\sum_{i=1}^{n_r}\sum_{k=1}^{K}\frac{a_i^kb_i^k\bar{N_s}}{(c\bar{N_r}\bar{N_s}+b_i^k)((c+a_i^k)\bar{N_r}\bar{N_s}+b_i^k)}\right)-
KP_1\left(\sum_{i=1}^{n_r}\sum_{k=1}^{K}\log\left(1+\frac{a_i^k}{\frac{b_i^k}{\bar{N_r}\bar{N_s}}+c}\right)\right)=0
\end{align}\normalsize
\end{figure*}

\section{Results and Discussion}

 In this section, we present numerical results to study the EE. The results are plotted for $\zeta=1$, $P_dL_d=5\times 10^{-6}$ W, $P_v/Q=5\times 10^{-4}$ W and $P_f=5$ W. Fig. \ref{fig_ee1} plots the EE versus the number of source antennas $N_s$ for different transmit powers $p_u$. The side lengths of the HMIMO surface at the BS are fixed at $L_{s,x}=L_{s,y}=5\lambda$, and  at each user are fixed at $L_{r,y}=L_{r,y}=1\lambda$. The value of $N_r$ is set as $N_r^*$ obtained using Theorem 2. The number of DoF at the source side can be computed as $n_s=77$. The figure shows that for $p_u=.001$ W and $p_u=.01$ W, the optimal number of BS antennas that maximizes the EE is $N_s^*=273$ and $N_s^*=127$ respectively. These values match with the solution obtained by solving \eqref{eq_implicitX} and \eqref{eq_implicity} in Theorem 2. For $p_u=1$ W, the number $N_s^*=n_s=77$ maximizes the EE, which also matches with the solution of Theorem 2. Interestingly, we see that for small values of $p_u$ which corresponds to the noise-limited scenario, it is beneficial in terms of EE to put more antennas than the available DoF in the HMIMO surface at the BS. On the other hand, for large $p_u$ values, which corresponds to the interference-limited scenarios, it suffices to just have as many antennas as the DoF created by scattering at transmit side to maximize the EE. This observation matches with our discussion after Theorem 1 that the impact of $N_s$ and $N_r$ on  achievable rates is dominant in noise-limited scenarios. 
 
 In Fig. \ref{fig_ee2} we plot the EE versus $N_s$ for $L_{s,x}=L_{s,y}=7\lambda$, and  $L_{r,y}=L_{r,y}=1\lambda$. We see that as the side lengths and consequently the number of DoF $n_s$ become larger, it becomes less beneficial to add more antennas beyond the maximum DoF $n_s$ that the BS HMIMO surface offers. Thus, the higher the values of $p_u$ are, and the larger the HMIMO surfaces are, the closer is the optimal number of transmit antennas to the number of DoF. Adding more antennas for these scenarios will just increase the power consumption without yielding any noticeable improvement in the achievable sum-rate. 
 
\begin{figure}[t]
\centering
\includegraphics[width=.85\columnwidth]{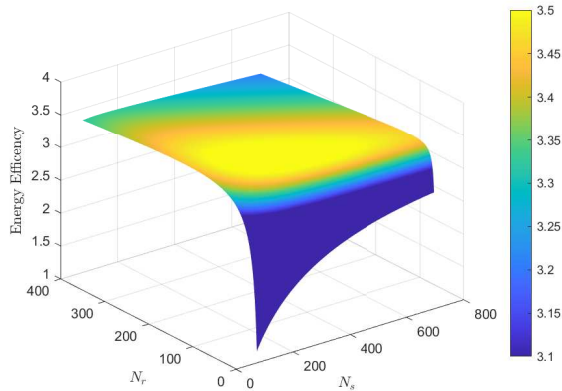}
\caption{EE versus the number of source and receive antennas $N_s$ and $N_r$ for $K=3$ users, $p_u=0.001$, $L_{s,x}=L_{s,y}=5\lambda$, and  $L_{r,x}=L_{r,y}=1\lambda$.}
\label{fig_ee3D}
\end{figure}

Fig. \ref{fig_ee3D} shows the EE against different numbers $N_s$ and $N_r$ of antennas placed within spatially-constrained HMIMO surfaces with side lengths $L_{s,x}=L_{s,y}=5\lambda$ and $L_{r,y}=L_{r,y}=1\lambda$ at the BS and user ends respectively. The values of $n_s$ and $n_r$ for these side lengths are $77$ and $3$. The figure shows that the optimal numbers of source and receive antennas are $N_s^*=272$ and $N_r^*=90$ respectively, which match with the solutions obtained by analytically solving \eqref{eq_implicitX} and \eqref{eq_implicity} as detailed in Thm. 2. For the considered surface size,  $(N_s^*, N_r^*)$ take values greater than $(n_s,n_r)$ as a noise-limited scenario is considered.

\section{conclusion}
In this paper, we derived the downlink ergodic achievable rate expression for a multi-user HMIMO communication system under MRT precoding, as a function of the side lengths of the HMIMO surfaces at the BS and users, and the number of antennas arranged in these surfaces. Utilizing the derived achievable sum-rate expression, an EE maximization problem was formulated to find the optimal numbers of source and receive antennas to be placed in the HMIMO surfaces, subject to constraints on these numbers to be larger than the DoFs offered by the channel. We derived an implicit solution for this problem. Also, we proved the solution to be globally optimal. The results revealed that in noise-limited scenarios, it is beneficial to have more antennas than the number of DoF to maximize the EE, while in the interference-limited scenarios, the optimal numbers of source and receive antennas is equal to the number of DoF offered by the HMIMO channel. This work could be extended for uplink multi-user HMIMO systems under different transmission schemes. 



\section*{APPENDIX}


The objective function of \textit{(P1)} in \eqref{eqn_maxee} has a strictly concave numerator and an affine denominator. To show that the numerator is strictly concave, we denote it by $ f(N_s,N_r)$ and check that its Hessian matrix is negative definite, i.e., $\nabla^2f(N_s,N_r)\prec0$ for $N_s>0, N_r>0$ as follows. The second-order derivative with respect to $N_s$ of the numerator $f(N_s,N_r)$ is given by $\frac{\partial ^2f(N_s,N_r)}{\partial N_s^2}=\sum_{i=1}^{n_r}\sum_{k=1}^{K}\frac{-a_ib_iN_r^2\left(\left(2c^2+2a_ic\right)N_rN_s+2b_ic+a_ib_i\right)}{\left(cN_rN_s+b_i\right)^2\left(\left(c+a_i\right)N_rN_s+b_i\right)^2}<0$,
where  the constants $a_i,b_i,c>0$ as they are functions of variances. Also, we can straightforwardly check that the determinant of the Hessian matrix is positive, i.e. $\text{det}(\nabla^2f)>0$, and the domain of the function forms a convex hull. Therefore, $\nabla^2f(N_s,N_r)\prec 0$ and the numerator of \textit{(P1)}, $f(N_s,N_r)$, is strictly concave. As a result, the objective function of \textit{(P1)}  in \eqref{eqn_maxee} is a strictly pseudo-concave function, being the ratio between a strictly concave function and an affine function. Strictly pseudo-concave functions are known either to  be  monotonically increasing or to admit a unique stationary point, which coincides with the function’s global maximizer. 
To see whether the objective function in \eqref{eqn_maxee} admits a stationary point, we observe its first-order derivatives with respect to $N_s$ and $N_r$ given on the left hand sides of  \eqref{eq_implicitX} and \eqref{eq_implicity} respectively ($N_s$ and $N_r$ are represented as $\bar{N_s}$ and $\bar{N_r}$ to facilitate the writing of Theorem 2). The first-order derivative  with respect to $N_s$ is positive as $N_s\rightarrow 0$ and it is negative as $N_s\rightarrow \infty$. A similar observation can be made about the first-order derivative with respect to $N_r$ in  \eqref{eq_implicity}. Therefore, the objective function has a unique stationary point $(\bar{N}_s, \bar{N}_r)$ in the range $N_s, N_r>0$. The solution in Theorem 2  either returns this stationary point or the values $(n_s, n_r)$ if $(\bar{N}_s, \bar{N}_r) \leq (n_s, n_r)$ under constraints \eqref{C1} and \eqref{C2}. Therefore, the solution in Thm. 2 is unique and global.


\vspace{-.05in}
\bibliographystyle{IEEEtran}
\bibliography{IEEEabrv,bibliography}

\begin{thebibliography}{10}
\providecommand{\url}[1]{#1}
\csname url@samestyle\endcsname
\providecommand{\newblock}{\relax}
\providecommand{\bibinfo}[2]{#2}
\providecommand{\BIBentrySTDinterwordspacing}{\spaceskip=0pt\relax}
\providecommand{\BIBentryALTinterwordstretchfactor}{4}
\providecommand{\BIBentryALTinterwordspacing}{\spaceskip=\fontdimen2\font plus
\BIBentryALTinterwordstretchfactor\fontdimen3\font minus
  \fontdimen4\font\relax}
\providecommand{\BIBforeignlanguage}[2]{{%
\expandafter\ifx\csname l@#1\endcsname\relax
\typeout{** WARNING: IEEEtran.bst: No hyphenation pattern has been}%
\typeout{** loaded for the language `#1'. Using the pattern for}%
\typeout{** the default language instead.}%
\else
\language=\csname l@#1\endcsname
\fi
#2}}
\providecommand{\BIBdecl}{\relax}
\BIBdecl

\bibitem{ref_mmimo}
E.~Björnson, L.~Sanguinetti, H.~Wymeersch, J.~Hoydis, and T.~L. Marzetta,
  ``Massive {MIMO} is a reality—what is next?: Five promising research
  directions for antenna arrays,'' \emph{Digit. Signal Process.}, vol.~94, pp.
  3--20, 2019.

\bibitem{ref_howmanyant}
J.~Hoydis, S.~ten Brink, and M.~Debbah, ``Massive mimo in the ul/dl of cellular
  networks: How many antennas do we need?'' \emph{IEEE J. on Sel. Areas in
  Commun.}, vol.~31, no.~2, pp. 160--171, 2013.

\bibitem{ref_mimogen}
E.~G. Larsson, O.~Edfors, F.~Tufvesson, and T.~L. Marzetta, ``Massive {MIMO}
  for next generation wireless systems,'' \emph{IEEE Commun. Mag.}, vol.~52,
  no.~2, pp. 186--195, 2014.

\bibitem{ref_channelmodel2}
A.~Pizzo, T.~L. Marzetta, and L.~Sanguinetti, ``Spatially-stationary model for
  holographic {MIMO} small-scale fading,'' \emph{IEEE J. on Sel. Areas in
  Commun.}, vol.~38, no.~9, pp. 1964--1979, 2020.

\bibitem{ref_surv50hmimo}
T.~{Gong} \emph{et~al.}, ``{Holographic {MIMO} Communications: Theoretical
  Foundations, Enabling Technologies, and Future Directions},'' \emph{arXiv
  e-prints}, p. arXiv:2212.01257, Dec. 2022.

\bibitem{ref_hmimochannelmodel}
A.~Pizzo, T.~Marzetta, and L.~Sanguinetti, ``Holographic {MIMO} communications
  under spatially stationary scattering,'' in \emph{Asilomar Conf. on Signals,
  Syst., and Computers}, 2020, pp. 702--706.

\bibitem{ref_hmimotrends}
C.~Huang, S.~Hu, G.~C. Alexandropoulos, A.~Zappone, C.~Yuen, R.~Zhang, M.~D.
  Renzo, and M.~Debbah, ``Holographic {MIMO} surfaces for 6g wireless networks:
  Opportunities, challenges, and trends,'' \emph{IEEE Wireless Commun.},
  vol.~27, no.~5, pp. 118--125, 2020.

\bibitem{ref_muhmimo}
L.~Wei \emph{et~al.}, ``Multi-user holographic {MIMO} surfaces: Channel
  modeling and spectral efficiency analysis,'' \emph{IEEE J. of Sel. Topics in
  Signal Process.}, vol.~16, no.~5, pp. 1112--1124, 2022.

\bibitem{ref_dofhmimo}
A.~Pizzo, T.~L. Marzetta, and L.~Sanguinetti, ``Degrees of freedom of
  holographic {MIMO} channels,'' in \emph{IEEE 21st Int. Workshop on Signal
  Process. Advances in Wireless Commun. (SPAWC)}, 2020, pp. 1--5.

\bibitem{ref_hmimocommodes}
N.~Decarli and D.~Dardari, ``Communication modes with large intelligent
  surfaces in the near field,'' \emph{IEEE Access}, vol.~9, pp.
  165\,648--165\,666, 2021.

\bibitem{ref_irselements}
Q.-U.-A. Nadeem, A.~Zappone, and A.~Chaaban, ``Intelligent reflecting surface
  enabled random rotations scheme for the miso broadcast channel,'' \emph{IEEE
  Trans. on Wireless Commun.}, vol.~20, no.~8, pp. 5226--5242, 2021.

\bibitem{ref_ach_rate}
J.~Jose, A.~Ashikhmin, T.~L. Marzetta, and S.~Vishwanath, ``Pilot contamination
  and precoding in multi-cell {TDD} systems,'' \emph{IEEE Trans. on Wireless
  Commun.}, vol.~10, no.~8, pp. 2640--2651, 2011.

\bibitem{ref_energy_hmimos}
S.~Zeng, H.~Zhang, B.~Di, H.~Qin, X.~Su, and L.~Song, ``Reconfigurable
  refractive surfaces: An energy-efficient way to holographic mimo,''
  \emph{IEEE Commun. Lett.}, vol.~26, no.~10, pp. 2490--2494, 2022.

\end{thebibliography}

\end{document}